\documentclass[conference]{IEEEtran}
\IEEEoverridecommandlockouts
\usepackage{algorithm}
\usepackage{xcolor,soul,framed} 
\usepackage{color, colortbl}
\usepackage[pdftex]{graphicx}
\usepackage[switch]{lineno}
\usepackage{array}
\usepackage{url}
\usepackage{tabularx}
\usepackage{lipsum,booktabs}
\usepackage{float}
\usepackage{threeparttable}
\usepackage{multirow}
\usepackage{makecell}
\usepackage{hyperref}
\usepackage[caption=false]{subfig}
\usepackage[T1]{fontenc}
\usepackage[noadjust]{cite}
\usepackage{longtable} 

\newcolumntype{L}[1]{>{\raggedright\arraybackslash}m{#1}}
\newcolumntype{R}[1]{>{\raggedleft\arraybackslash}m{#1}}
\newcolumntype{C}[1]{>{\centering\arraybackslash}m{#1}}
\newcommand{\tvar}[2]{$\textbf{#1}_\textbf{#2}$}

\definecolor{LightCyan}{rgb}{0.88,1,1}
\definecolor{Gray}{gray}{0.9}
\definecolor{Gray2}{gray}{0.96}

\newcommand*{\belowrulesepcolor}[1]{%
  \noalign{%
    \kern-\belowrulesep 
    \begingroup 
      \color{#1}%
      \hrule height\belowrulesep 
    \endgroup 
  }%
} 
\newcommand*{\aboverulesepcolor}[1]{%
  \noalign{%
    \begingroup 
      \color{#1}%
      \hrule height\aboverulesep 
    \endgroup 
    \kern-\aboverulesep 
  }%
} 

\usepackage{cite}
\usepackage{amsmath,amssymb,amsfonts}
\usepackage{algorithmic}
\usepackage{textcomp}
\def\BibTeX{{\rm B\kern-.05em{\sc i\kern-.025em b}\kern-.08em
    T\kern-.1667em\lower.7ex\hbox{E}\kern-.125emX}}

\DeclareRobustCommand*{\IEEEauthorrefmark}[1]{%
    \raisebox{0pt}[0pt][0pt]{\textsuperscript{\footnotesize\ensuremath{#1}}}}
    
\begin{document}

\title{An AI-driven EDA Algorithm-Empowered VCO and LDO Co-Design Method \\
\thanks{The work of the first author was supported by the Engineering and Physical Sciences Research Council (EPSRC) under Ph.D. grant.}
}

\author{
\IEEEauthorblockN{
Yijia Hao\IEEEauthorrefmark{1},
Maarten Strackx\IEEEauthorrefmark{2},
Miguel Gandara\IEEEauthorrefmark{3},
Sandy Cochran\IEEEauthorrefmark{1},
Bo Liu\IEEEauthorrefmark{1}}
\IEEEauthorblockA{\IEEEauthorrefmark{1}University of Glasgow, UK}
\IEEEauthorblockA{\IEEEauthorrefmark{2}Magics Technologies, Belgium}
\IEEEauthorblockA{\IEEEauthorrefmark{3}Mediatek, USA}
\IEEEauthorblockA{2357677H@student.gla.ac.uk, Bo.Liu@glasgow.ac.uk}
}

\maketitle

\begin{abstract}
Traditionally, the output noise and power supply rejection of low-dropout regulators (LDOs) are optimized to minimize power supply fluctuations, reducing their impact on the low-frequency noise of target voltage-controlled oscillators (VCOs). However, this sequential design approach does not fully address the trade-offs between high-frequency and LDO-induced low-frequency phase noise. To overcome this limitation, this paper presents a co-design method for low phase-noise LC-tank VCOs powered by LDOs. It is difficult to carry out the co-design using traditional manual design techniques. Hence, an efficient AI-driven EDA algorithm is used. To validate the proposed method, a 5.6 GHz LC-tank VCO with an integrated LDO is designed using a 65 nm CMOS process. Simulations show that the co-design method improves phase noise by 1.2 dB at a 1 MHz offset and reduces dynamic power consumption by 28.8\%, with FoM increased by 2.4 dBc/Hz compared to the conventional sequential design method.
\end{abstract}


\section{Introduction}
Voltage-controlled oscillators (VCOs) are crucial components in high performance systems, particularly in wireless communication and radio-frequency applications. One of the major challenges in designing VCOs is optimizing phase noise (PN), which can lead to degraded signal purity and reduced system performance \cite{4684621}. While traditional approaches have focused on improving the inherent noise performance of the oscillator's core components, such as minimizing thermal and flicker noise in the transistors or enhancing the quality factor of the LC tank, external influences also play a significant role. Power supply noise can introduce additional PN through power supply rejection (PSR) path and VCO's frequency pushing factor \cite{huang15324GHzADPLL2014}. To mitigate this impact, low-dropout regulators (LDOs) are commonly used \cite{ursoAnalysisDesignPower2020}. However, the LDO itself introduces new challenges, as its low-frequency noise can be up-converted into PN, complicating the overall noise management strategy in VCO designs.

To improve the PN, a natural idea is to design LDO and VCO together. \cite{xuejinwangSystematicDesignSupply2008} carries out a comprehensive analysis between frequency pushing and power supply-induced PN. Based on the influence of the combined effect of the LDO’s PSR and the VCO’s inherent phase noise sensitivity to supply noise, the guidelines of LDO design are obtained. Reducing the width of the VCO’s switching device is suggested in \cite{xuejinwangSystematicDesignSupply2008}, which mitigates the frequency pushing effect. However, this may increase thermal noise, potentially degrading the VCO’s PN performance at higher frequencies. Therefore, a holistic approach to obtain the optimal trade-off considering various kinds of noise is needed to obtain the truly optimal design. 

\begin{figure}[t]
\centering
\includegraphics[width=3.3in]{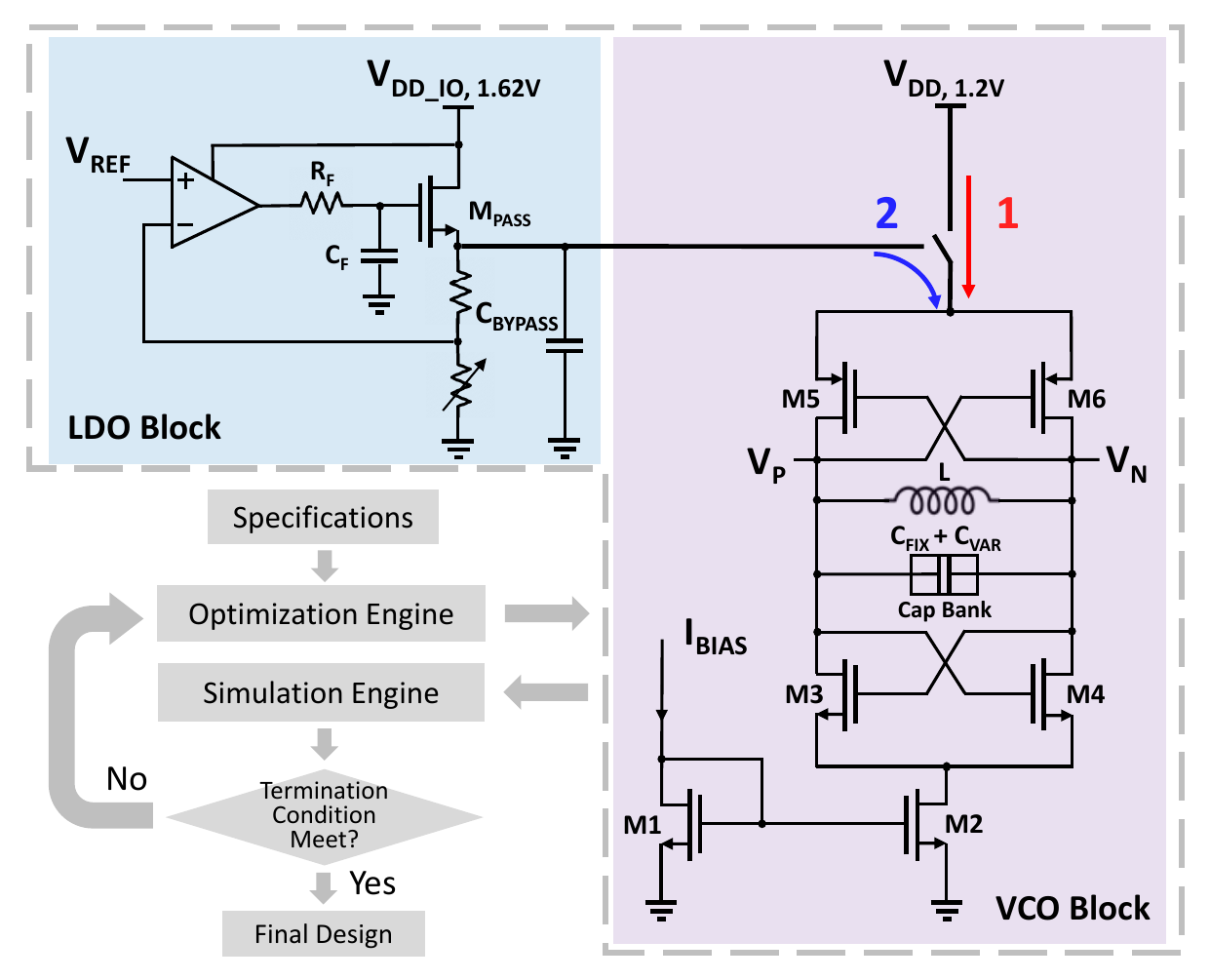}
\caption{Abstract of VCO and LDO co-design method including the schematic diagram of the LC-tank VCO with an integrated LDO. Two design approaches: the sequential approach, which involves two distinct design phases, and the co-design approach, which optimizes both building blocks simultaneously.}
\label{ldovco}
\end{figure}

For the traditional manual design method, it can be challenging to derive formulas due to the complex trade-offs described above. Hence, this paper presents an LDO and VCO co-design method empowered by an AI-driven EDA algorithm called efficient surrogate model-assisted sizing method for high-performance analog building blocks (ESSAB) \cite{budakEfficientAnalogCircuit2022}. The method optimizes the PN of an LC-tank VCO and the PSR of an integrated LDO while also accounting for PVT corners. Notably, while most existing AI-driven design research focuses on individual analog building blocks, this work emphasizes subsystem-level co-design, offering a more holistic and effective optimization approach.

To validate the proposed co-design method, a 5.6 GHz LC-tank VCO, regulated by an LDO, is designed using a 65 nm CMOS process. Simulation results show that the co-design approach improves the VCO's phase noise by 1.2 dB at a 1 MHz offset and reduces dynamic power consumption by 28.8\%, with FoM increased by 2.4 dBc/Hz compared to the traditional sequential design approach using the same AI-driven EDA algorithm.  

The paper is structured as follows: Section II presents the building blocks for the targeted LDO-VCO. Section III explains the implementation details of co-design method. Section IV compares the designs obtained by the traditional sequential method and the co-design method. Conclusions are presented in Section V.

\section{Architecture of LDO-VCO}
A diagram of an LDO-regulated VCO is shown in Fig. \ref{ldovco}, where a cross-coupled LC-tank VCO is adopted \cite{soltanianUltraCompactDifferentiallyTuned2007}. The LC tank comprises an inductor, capacitor, and a varactor array. In parallel, the cross-coupled transistor pairs produce negative resistance to counteract the losses present in the LC tank. The LDO on the top provides the load current needed for the VCO. It consists of an error amplifier, a low-pass filter, an NMOS pass transistor, and a feedback divider. The error amplifier is implemented using a two-stage Miller-compensated op-amp. In addition, a bypass capacitor is used at the LDO output. 

\section{AI-Driven Co-Design Method}
\subsection{Setup of the Sizing Problem}
\subsubsection{Design Variables}
Table \ref{variable} details the ranges of the 43 design variables, where $W$, $R$, $NT$, $S$, $GR$ denotes the width, inner radius, number of turns, spacing between conductors and guard ring width of the inductor; $L$, $W$, $F$, $M$ represent the channel length, width per finger, number of fingers and number of multiplier of the transistors; $N_H$, $N_V$, and $M_{bot}$ represent the number of horizontal fingers, vertical fingers, and bottom starting layer for the MOM capacitors in the VCO. The inductor and MOM capacitor are implemented using PDK components. The biasing circuits for VCO and LDO, the LDO feedback divider, the varactor and the bypass capacitor are maintained constant.

\begin{table}[htb]
\scriptsize
\caption{Design variables and search ranges of the CMOS cross-coupled LC oscillator and the LDO.}
\label{variable}
\renewcommand{\arraystretch}{1.23}
\centering
\begin{tabular}{p{0.6cm} p{0.8cm} p{0.9cm} p{0.9cm} p{0.9cm} p{0.9cm} p{0.9cm}}
\toprule
\textbf{ } & \textbf{Var.} & \textbf{Unit} & \textbf{Lower bound} & \textbf{Upper bound} & \textbf{Co-design} & \textbf{Se-design} \\
\midrule
 \multirow{16}{*}{\textbf{VCO}} 
 & \tvar{M}{2} & integer & 1 & 1000 & 300 & 872\\ \cline{2-7}
 & \tvar{L}{3,4} & m & 60n & 240n & 225n & 239n\\ 
 & \tvar{W}{3,4} & m & 1u & 6u & 1.22u & 4.60u\\ 
 & \tvar{F}{3,4} & integer & 2 & 32 & 7 & 10\\ 
 & \tvar{M}{3,4} & integer & 1 & 10 & 8 & 2\\ \cline{2-7}
 & \tvar{L}{5,6} & m & 60n & 240n & 205n & 75n\\
 & \tvar{W}{5,6} & m & 1u & 6u & 1.96u & 1.72u\\
 & \tvar{F}{5,6} & integer & 2 & 32 & 11 & 13\\ 
 & \tvar{M}{5,6} & integer & 1 & 10 & 6 & 10\\ \cline{2-7}
 & \tvar{N}{H} & integer & 10 & 200 & 74 & 94\\ 
 & \tvar{N}{V} & integer & 10 & 200 & 95 & 88\\ 
 & \tvar{M}{bot} & integer & 1 & 3 & 1 & 1\\ \cline{2-7}
 & \tvar{W}{} & m & 3u & 30u & 28.2u & 27.9u\\ 
 & \tvar{R}{} & m & 15u & 90u & 89.4u & 76.6u \\ 
 & \tvar{N}{} & integer & 1 & 3 & 1 & 1\\ 
 & \tvar{S}{} & m & 2u & 4u & 2.67u & 3.18u\\ 
 & \tvar{GR}{} & m & 10u & 40u & 28.7u & 21.7u\\ \hline \hline
\multirow{24}{*}{\textbf{LDO}} 
 & \tvar{L}{nLoad} & m& 500n & 10u & 6.64u & 8.28u\\
 & \tvar{W}{nLoad} & m & 400n & 10u & 3.93u & 500n\\ 
 & \tvar{F}{nLoad} & integer & 2 & 32 & 25 & 3\\ 
 & \tvar{M}{nLoad} & integer & 1 & 10 & 2 & 2\\ \cline{2-7}
 & \tvar{L}{pIn} & m & 400n & 10u & 5.95u & 470n\\ 
 & \tvar{W}{pIn} & m & 400n & 10u & 3.25 & 1.43\\ 
 & \tvar{F}{pIn} & integer & 2 & 32 & 29 & 5\\ 
 & \tvar{M}{pIn} & integer & 1 & 10 & 5 & 1\\ \cline{2-7}
 & \tvar{L}{bias} & m & 400n & 10u & 5.55u & 3.63u\\ 
 & \tvar{W}{bias} & m & 400n & 10u & 4.58u & 9.14u\\ 
 & \tvar{F}{bias} & integer & 2 & 32 & 14 & 22\\ 
 & \tvar{M}{bias} & integer & 1 & 10 & 7 & 9\\ 
 & \tvar{M}{biasIn} & integer & 1 & 10 & 5 & 6\\ 
 & \tvar{M}{biasOut} & integer & 1 & 10 & 8 & 1\\ \cline{2-7}
 & \tvar{L}{nOut} & m & 500n & 10u & 2.53u & 3.42u\\ 
 & \tvar{W}{nOut} & m & 400n & 10u & 2.08u & 6.35u\\ 
 & \tvar{F}{nOut} & integer & 2 & 32 & 31 & 20\\ 
 & \tvar{M}{nOut} & integer & 1 & 10 & 7 & 7\\ \cline{2-7}
 & \tvar{C}{C} & F & 1p & 100p & 67p & 60p\\ 
 & \tvar{R}{C} & ohm & 1 & 1M & 989K & 514K\\ \cline{2-7}
 & \tvar{C}{F} & F & 1p & 200p & 182p & 156p\\ 
 & \tvar{R}{F} & ohm & 1 & 2M & 1.66M & 1.17M\\ \cline{2-7}
 & \tvar{L}{pass} & m & 1.2u & 10u & 1.69u & 1.62u\\ 
 & \tvar{W}{pass} & m & 500n & 10u & 8.86u & 5.96u\\
 & \tvar{F}{pass} & integer & 2 & 100 & 47 & 35\\ 
 & \tvar{M}{pass} & integer & 1 & 32 & 15 & 15\\ 
\bottomrule
\end{tabular}
\vspace{-10pt}
\end{table}

\subsubsection{Testbench and Measures}
A 65 nm CMOS process is used. The VCO varactor is set to work at its highest frequency where supply pushing is the highest. The performance metrics include oscillation frequency $f_0$, phase noise $PN(\Delta f)$ at frequency offsets $\Delta f$ of 100 kHz, 1 MHz and 10 MHz, total power consumption $P_{dyn}$, and the FoM at 1 MHz frequency offset \cite{tsangHighFigureMerit2003}:
\begin{equation}\label{EQ5}
FoM = -10\text{log} [(\frac{\Delta f}{f_0})^2\cdot \frac{P_{dyn}}{1mW}] - PN(\Delta f),
\end{equation}
Additionally, the maximum PSR, phase margin (PM), and maximum $V_{DD}$ are extracted. The process corners considered include fast NMOS/fast PMOS, fast NMOS/slow PMOS, slow NMOS/slow PMOS, and slow NMOS/fast PMOS  in combination with min inductor/max inductor and min capacitor/max capacitor. $-$55°C and 125°C are considered as temperature corners. And the lowest supply voltage $V_{DD\_IO}$ (1.8 V $\cdot$ 90\%) is used. In total, 32 corners are considered.

\subsubsection{Objective and Constraints}
The targeting oscillation frequency is 5.5 GHz with a power consumption less than 7 mW. The phase noise constraints at 100 kHz, 1 MHz and 10 MHz are set to be $-$94 dBc/Hz, $-$120 dBc/Hz, and $-$140 dBc/Hz respectively, which meet industrial standards. The optimization objective for both approaches is FoM. The remaining performance parameters are set as constraints. These apply to all 32 corners.

\subsection{Sizing Flow and Considerations}
For the sequential design method, the VCO is first optimized independently with an ideal 1.2 V power supply to achieve an optimal FoM, as illustrated in Fig. \ref{ldovco}. The LDO is then incorporated and optimized for this VCO load to generate a clean supply. Although this approach seems intuitive, the primary challenge lies in preventing the LDO noise from being upconverted and affecting the PN of the VCO. To mitigate these, the VCO may need to be re-designed to reduce its sensitivity to the noise introduced by the LDO. This iterative process often requires multiple sizing loops, which can be time-consuming and labor-intensive. Additionally, a bypass capacitor is typically used to reduce high-frequency supply noise, but its impact is often overlooked in the sizing stage of VCO. Not considering this can lead to discrepancies between the designed and measured PN performance due to changed VCO’s voltage swing.

To address above issues, a simultaneous co-design of the LDO and VCO is implemented. By treating them as an integrated system rather than separate blocks, mutual interaction is considered throughout the design process, ensuring that the contributors are jointly optimized to minimize PN. Fig. \ref{ldovco} presents the whole sizing flow. 

\subsection{Sizing Algorithm}
The ESSAB algorithm based on a single objective Bayesian optimization framework is used for both the sequential and co-design methods for a apple-to-apple comparison. It starts by initializing a database and iteratively refining designs until a predefined stopping criterion is met. Each iteration involves selecting the top candidate and applying differential evolution operations. An online ANN model and beta ranking are used to guide the selection of the most promising candidate for simulation. The steps are abstracted as Algorithm \ref{alg:optimization} and algorithm details can be found in \cite{budakEfficientAnalogCircuit2022}.



    
    
    
    
    



\begin{algorithm}
\small
\caption{Optimization Framework}
\label{alg:optimization}

\begin{algorithmic}[1]
\STATE Initialize Database
\STATE Finish $\gets$ false

\WHILE{Finish = false}
    \STATE Rank and select top $\lambda$ designs
    \STATE Apply Differential Evolution (DE) operations
    \STATE Train ANN and predict performance
    \STATE Select best predicted design and simulate
    \IF{Stopping criterion met} \STATE Finish $\gets$ true
    \ELSE \STATE Update Database
    \ENDIF
\ENDWHILE

\STATE Output Final Design

\end{algorithmic}
\end{algorithm}

\section{Pre-layout Sizing Results and Analysis}
Using Cadence Virtuoso and ESSAB tool implemented in MATLAB on a workstation with an AMD Ryzen Threadripper PRO 3975WX (32 cores, 290 GB RAM), the sequential method used 18 hours in total with 7 hours in VCO sizing and 11 hours in LDO sizing, while the co-design method used 6 hours in total. The sizing details for the obtained designs are provided in Table \ref{ldovco}, labeled as co-design and se-design, respectively. The pre-layout simulation results are summarized in Table \ref{result vco}. The results of nominal corner and the corner with slow NMOS/slow PMOS, max inductor and max capacitor at 125°C (worst corner) are listed. 

To assess the impact of LDO in two approaches, the LDO output noise and PSR are analyzed. In addition, the VCO phase noise is extracted under two conditions: 1) powered by a 1.2 V ideal voltage source, and 2) powered by the LDO with a 1.62 V DC input and a 1.2 V output voltage. 

\begin{figure}[t]
\centering
\includegraphics[width=3.3in]{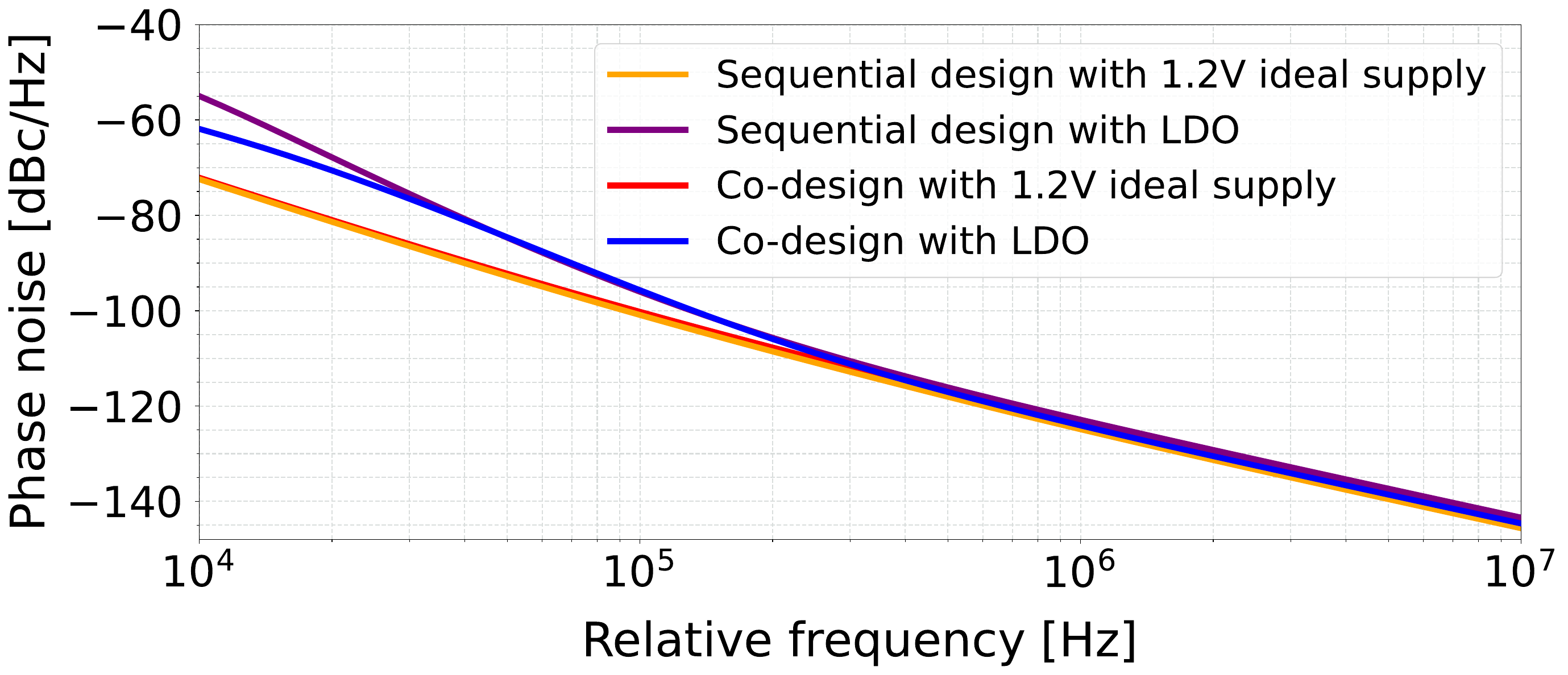}
\caption{Phase noise performance of the VCO designs with 1.2 V ideal supply and with LDO.}
\label{ldo}
\end{figure}

At a 100 kHz offset, the LDO noise significantly deteriorates the VCO phase noise. The primary contributor to the LDO noise differs between the two designs: For the co-designed LDO, the main source of noise is the thermal noise from the NMOS load in the input stage. For the sequential design, the primary contributor is the thermal noise from the PMOS input pair. In the PMOS input pair, the transconductance ($g_m$) is 61.89 $\mu S$ for the sequential design and 73.42 $\mu S$ for the co-designed LDO. For the NMOS load, $g_m$ is 73.9 $\mu S$ in the co-designed case, compared to 18.7 $\mu S$ for the sequential design, resulting in higher output noise for the co-designed LDO. Further, as shown in Fig. \ref{ldo}, the co-designed VCO with ideal supply has a slightly worse PN at 100 kHz due to the smaller switching transistors. With LDO's output noise, the co-designed system exhibits worse phase noise performance at the 100 kHz offset, with $-$95.6 dB compared to $-$96.2 dB. 











\begin{table}[t]
\scriptsize
\setlength{\tabcolsep}{2pt}
\caption{Specifications and pre-layout simulation results of the sequentially and co-designed LDO-VCO.}
\label{result vco}
\renewcommand{\arraystretch}{1.2}
\centering
\begin{tabular}{L{2.35cm} L{1cm} C{1.1cm} C{1.1cm} C{1.1cm} C{1.1cm}}
\toprule
\textbf{Symbol} & \textbf{Specs.} & \textbf{Se-design (Nominal)} & \textbf{Co-design (Nominal)} & \textbf{Se-design (Worst corner)} & \textbf{Co-design (Worst corner)} \\
\midrule
FoM (dBc/Hz) & Minimize & $-$190 & $-$192.4 & $-$187.3 & $-$187.8 \\
Frequency (GHz) & $\geq$ 5 & 5.69 & 5.60 & 5.35 & 5.27 \\
PN@100kHz (dBc/Hz) & $\leq$ $-$94 & $-$96.2 & $-$95.6 & $-$93.8 & $-$92.0 \\
PN@1MHz (dBc/Hz) & $\leq$ $-$120 & $-$122.9 & $-$124.1 & $-$120.9 & $-$119.7 \\
PN@10MHz (dBc/Hz) & $\leq$ $-$140 & $-$143.4 & $-$144.7 & $-$142 & $-$141.5 \\
P\textsubscript{dyn} (mW) & $\leq$ 7 & 6.40 & 4.56 & 6.60 & 4.33 \\
PSR\textsubscript{max} (dB) & $\leq$ $-$30 & $-$33.7 & $-$31.4 & $-$31.6 & $-$31.0 \\
V\textsubscript{DD,max} (V) & $\leq$ 1.32 & 1.24 & 1.23 & 1.24 & 1.23 \\
PM ($^\circ$) & $\geq$ 50 & 67 & 82 & 59 & 81 \\
\bottomrule
\end{tabular}
\end{table}


At 1 MHz offset, the primary contributor to phase noise is the thermal noise from the switching transistors, while the flicker noise of transistors M3-M6 dominates across corners. With ideal supply, M5 and M6 have five times smaller $W/L$ in the co-designed VCO, which reduces the effective transconductance of the switching pairs, resulting in higher thermal noise and 1 dB worse PN than se-design. However, with LDO incorporated, the PN performance of the se-design degrades from 124.9 dBc/Hz to 122.9 dBc/Hz, partly due to frequency pushing mechanism, which is significantly suppressed in the co-design. Consequently, the co-design achieves a 1.2 dB PN reduction at a 1 MHz offset and a 1.3 dB improvement at 10 MHz with the LDO incorporated. Additionally, the power consumption is reduced due to the lowered total capacitance at the output nodes. The decrease in transconductance reduces the current flowing to the LC tank and affects the startup time. However, the startup is still achieved reliably across corners thanks to the corner analysis during optimization. 

\begin{figure}[t] 
\centering
\subfloat[\label{3D}]{%
\includegraphics[width=0.85\linewidth]{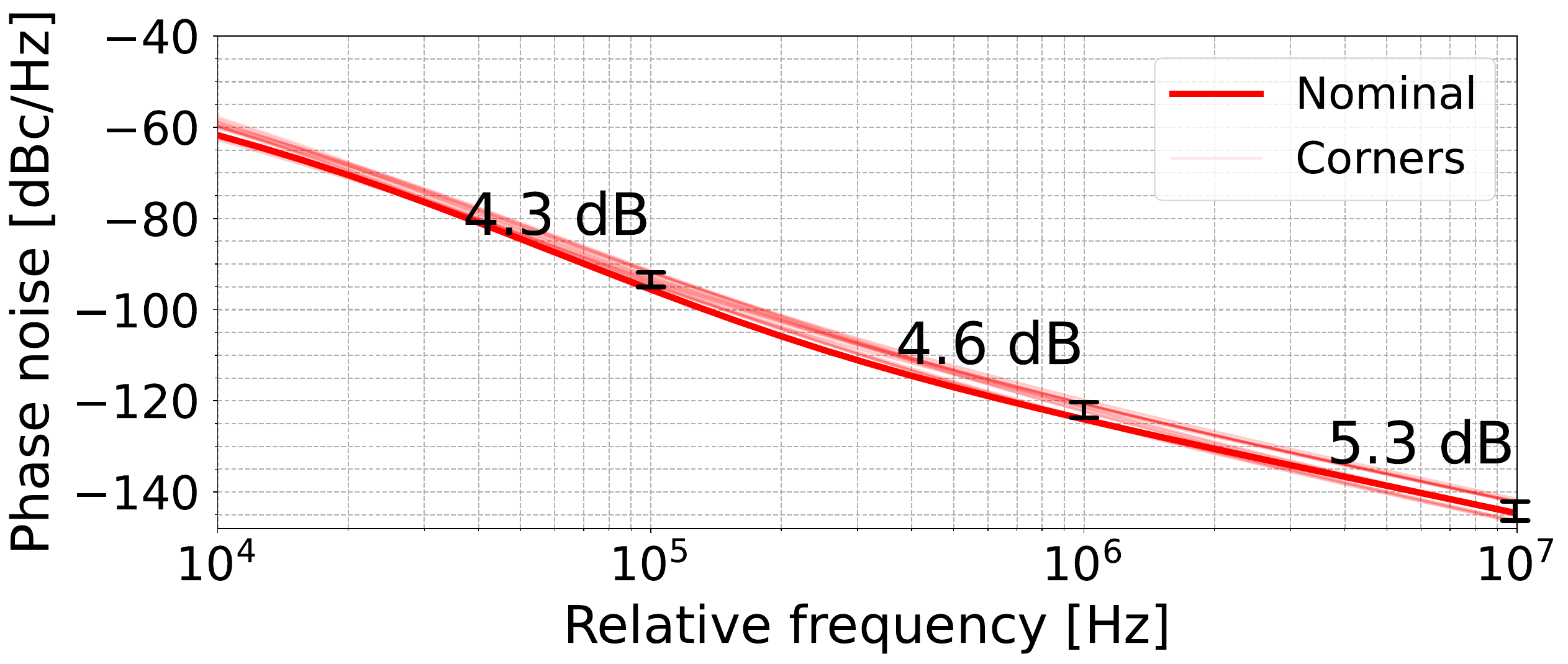}}
\vspace{-6pt}
\hfill
\subfloat[\label{2D}]{%
\includegraphics[width=0.85\linewidth]{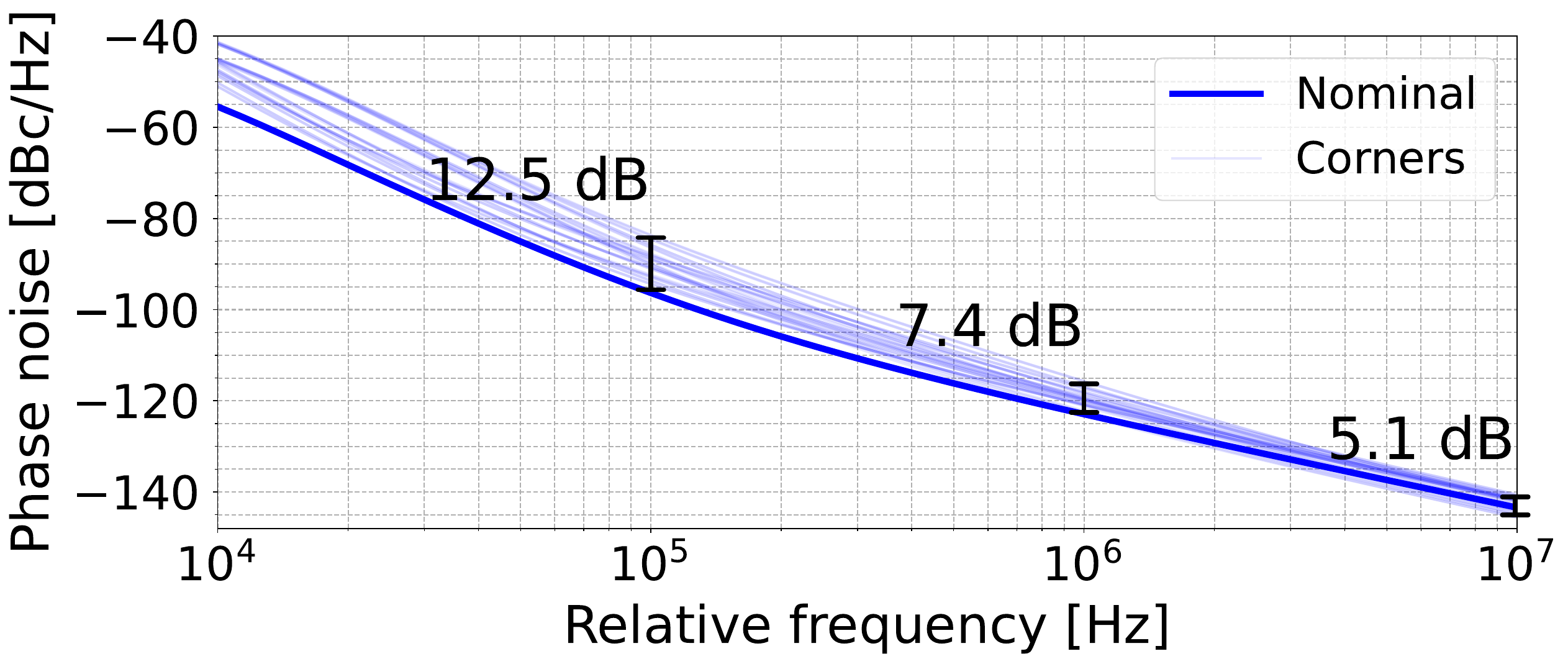}}
\caption{(a) Corner spread of PN for the co-designed LDO-VCO. (b) Corner spread of PN for the sequentially designed LDO-VCO. }
\label{corners} 
\end{figure}

In the sequentially designed LDO-VCO, the effect of the bypass capacitor is not well accounted for in the VCO design. To maintain low phase noise, the supply voltage swing is kept large. However, when a bypass capacitor is added at the $V_{DD}$ of the VCO, it flattens the voltage swing, which worsens the optimized phase noise performance. In contrast, the co-design approach considers the impact of the decoupling capacitor from the start. This approach constrains the reliance on voltage swings, resulting in lower phase noise. The inclusion of a bypass capacitor also degrades the VCO phase noise across corners. According to Fig. \ref{corners}, the pre-designed VCO exhibits a much larger variation in phase noise 
when the LDO and bypass capacitor are included. In contrast, the co-designed VCO shows a more consistent performance. Overall, the co-design approach leads to a better FoM of 2.4 dBc/Hz and superior performance across corners by accounting for block interactions and multiple effects. 

\begin{figure}[t]
\centering
\includegraphics[width=3.1in]{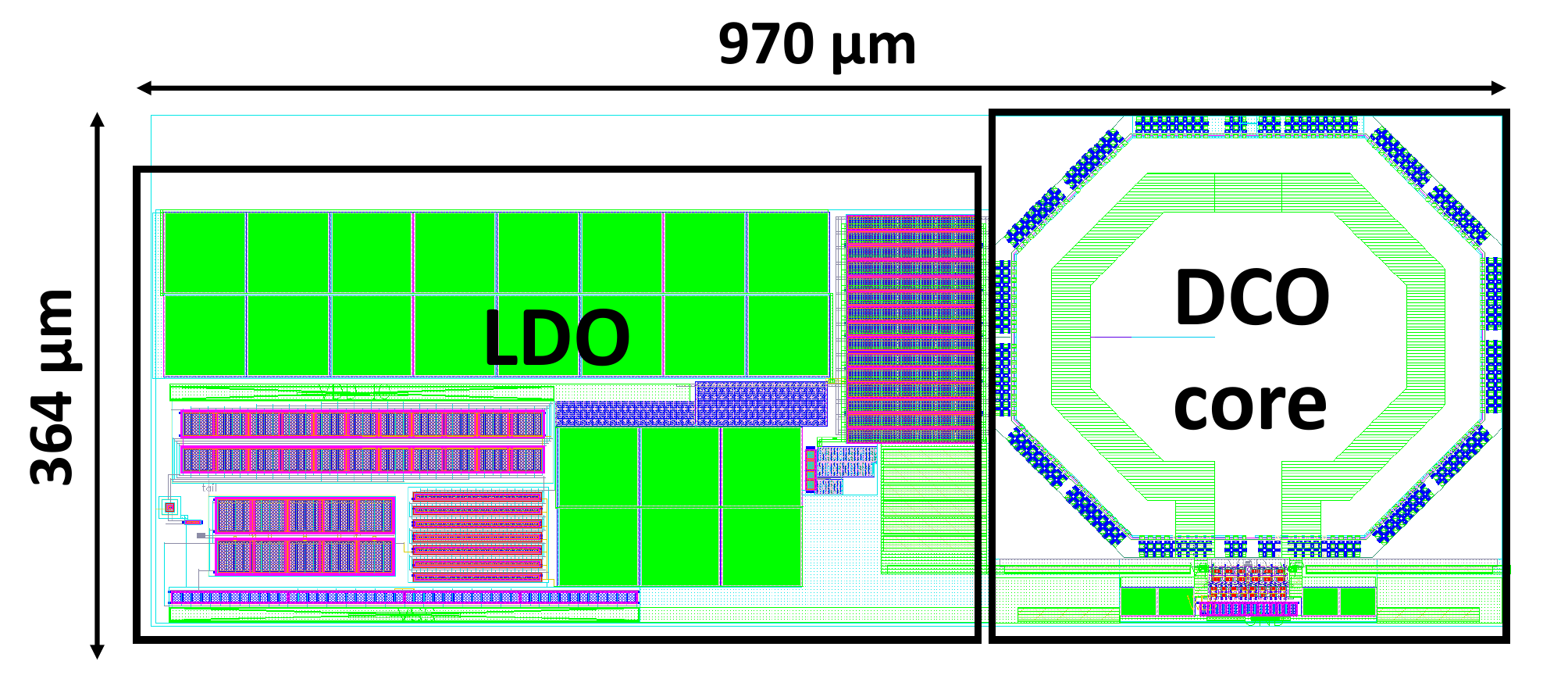}
\caption{Co-designed LDO-VCO layout.}
\vspace{-2pt}
\label{layout}
\end{figure}

The layout was manually implemented, shown in Fig. \ref{layout}. The post-layout simulation results are presented in Table \ref{post}. The impact of layout on the FoM is limited to 0.3 dB, which is caused by the routing parasitics in the $V_P$ and $V_N$ nets.

\begin{table}[h!]
\scriptsize
\setlength{\tabcolsep}{2pt}
\caption{Specifications and post-layout simulation results of the co-designed LDO-DCO.}
\label{post}
\renewcommand{\arraystretch}{1.2}
\centering
\begin{tabular}{L{3.2cm} L{1.6cm} C{1.6cm} C{1.6cm}}
\toprule
\textbf{Symbol} & \textbf{Specs.} & \textbf{Co-design (Nominal)} & \textbf{Co-design (Worst corner)}\\
\midrule
FoM (dBc/Hz) & Minimize & $-$192.1 & $-$187.8\\
Frequency (GHz) & $\geq$ 5 & 5.51 & 5.19\\
PN@100kHz (dBc/Hz) & $\leq$ $-$94 & $-$95.9 & $-$96.8\\
PN@1MHz (dBc/Hz) & $\leq$ $-$120 & $-$123.9 & $-$119.1\\
PN@10MHz (dBc/Hz) & $\leq$ $-$140 & $-$144.3 & $-$139.3\\
P\textsubscript{dyn} (mW) & $\leq$ 7 & 4.67 & 3.59\\
PSR (dB) & $\leq$ $-$30 & $-$30.2 & $-$30.1\\
V\textsubscript{DD,max} (V) & $\leq$ 1.32 & 1.22 & 1.22\\
PM ($^\circ$) & $\geq$ 50 & 82 & 81\\
\bottomrule
\end{tabular}
\vspace{-10pt}
\end{table}

\section{Conclusion}
This paper has presented a VCO and LDO co-design method for PN reduction and performance enhancement, which is empowered by ESSAB. Compared to the traditional sequential design method, the proposed approach improves the FoM by 2.4 dB, highlighting its ability to consider analog building block interactions and trade-offs to optimize overall performance.


\bibliographystyle{IEEEtran}
\bibliography{LDO-DCO,VCO}

\end{document}